\documentclass[journal]{IEEEtran}
\usepackage{xcolor,soul,framed} 
\usepackage[cmex10]{amsmath}
\usepackage{array}
\usepackage{mdwmath}
\usepackage{mdwtab}
\usepackage{eqparbox}
\usepackage{url}
\usepackage{bm}
\usepackage{mathtools}
\usepackage{amsfonts}
\usepackage{amssymb}	
\usepackage{slashed}    
\usepackage{dsfont}
\usepackage{graphicx}
\usepackage{subfigure}
\usepackage{booktabs}
\usepackage{multirow}
\usepackage{hyperref}
\hypersetup{breaklinks,colorlinks,citecolor=blue}
\usepackage{cite}
\usepackage{algorithm}
\usepackage{setspace}
\usepackage[noend]{algpseudocode}
\usepackage{tikz}
\usetikzlibrary{spy}
\makeatletter
\def\BState{\State\hskip-\ALG@thistlm}
\makeatother

\DeclareMathOperator*{\argminT}{argmin}
\DeclareMathOperator*{\argmaxT}{argmax}

\DeclareMathOperator*{\maxT}{max}
\usepackage{mathtools}
\DeclarePairedDelimiterX{\norm}[1]{\lVert}{\rVert}{#1}
\DeclarePairedDelimiter\abs{\lvert}{\rvert}%

\newcommand{\R}{\mathcal{R}}

\DeclarePairedDelimiterX{\snorm}[1]{{}_{\mathbb{S}^2}\!\lVert}{\rVert}{#1}
\DeclarePairedDelimiterX{\pnorm}[1]{{}_{\Psi}\!\lVert}{\rVert}{#1}

\begin{document}
\bstctlcite{IEEEexample:TIP}
\title{Sparse image reconstruction on the sphere:\\ a general approach with uncertainty quantification}
\author{Matthew~A.~Price, Luke~Pratley, Jason~D.~McEwen}

\markboth{IEEE TRANSACTIONS ON IMAGE PROCESSING
}{Price \MakeLowercase{\textit{et al.}}: Spherical optimization with uncertainty quantification}

\maketitle

\begin{abstract}

  Inverse problems defined naturally on the sphere are becoming increasingly of interest. In this article we provide a general framework for evaluation of
  inverse problems on the sphere, with a strong emphasis on flexibility and scalability. We
  consider flexibility with respect to the prior selection (regularization), the problem definition --- specifically the problem formulation (constrained/unconstrained) and
  problem setting (analysis/synthesis) --- and optimization adopted to solve the problem. We discuss and quantify the trade-offs
  between problem formulation and setting. Crucially, we consider the Bayesian interpretation
  of the unconstrained problem which, combined with recent developments in probability density
  theory, permits rapid, statistically principled uncertainty quantification (UQ) in the spherical setting.
  Linearity is exploited to significantly increase the computational efficiency of such UQ techniques,
  which in some cases are shown to permit analytic solutions.
  We showcase this reconstruction framework and UQ techniques on a variety of spherical inverse problems. The code discussed throughout is provided
  under a GNU general public license, in both C++ and Python.
\end{abstract}

\begin{IEEEkeywords}
  harmonic analysis, sampling, spheres, spherical wavelets, uncertainty quantification
\end{IEEEkeywords}

\IEEEpeerreviewmaketitle


\section{Introduction}

\IEEEPARstart{I}{ncreasingly} often one wishes to solve inverse problems natively on the sphere
($\mathbb{S}^2)$ rather than on $n$ dimensional Euclidean space ($\mathbb{R}^n)$, \textit{e.g.}
in astronomy and astrophysics \cite{M4, Wallis2017MASSMAPPY, Planck2018I},
biomedical imaging \cite{mcewen:s2let_ridgelets, tuch2004q}, and geophysics
\cite{ritsema2011s40rts}. Straightforwardly from Gauss' \emph{Theorema Egregium} --- which states that the curvature of surfaces embedded in $\mathbb{R}^3$ is immutable, and thus planar projections of curved manifolds (\emph{e.g.} the sphere) inherently incur (significant) distortions --- analysis over such domains
must necessarily be conducted natively on the sphere. Though many Euclidean techniques may provide
inspiration for counter-parts on the sphere, there are a still a great many
critical differences between these paradigms which must be considered. Typically, inverse
problems of interest, particularly on the sphere, are (often severely) ill-posed
and/or ill-conditioned, motivating the injection of prior knowledge to stabilize the
reconstruction. Such problems can be solved in a variety of ways (\emph{e.g.} sampling
methods and machine learning methods) though, for robustness and scalability, in the
spherical setting variational methods (\emph{e.g.} optimization) are the most
effective.

Due to recent advances in the theory of compressed sensing \cite{candes2006robust,
  candes2006compressive, Donoho2006} sparsity priors (\emph{e.g.} $\ell_1$-regularization)
are now routinely adopted, where the solution to an inverse problem can be constrained
and found by promoting sparsity in a dictionary, such as wavelets or gradient space (variational norms).
Recent developments in proximal convex optimization algorithms facilitate the practical application of non-differentiable priors, where they
can be distributed and scale to high dimensional parameter spaces \cite{Pratley2019widefield,Onose2016SOPT}.
The spherical counterparts for discrete gradient spaces \cite{McEwen2013Sparse},
wavelet families \cite{narcowich2006localized, baldi2009asymptotics, Starck2006Wavelets},
and scale-discretized wavelet families \cite{Wiaux2008WaveletSynthesis,
  Leistedt2013S2LET, mcewen:2013:waveletsxv, McEwen2015SpinWavelets,
  McEwenS2DWLocalisation} have been developed, and have found wide applications --- see
previous papers in this series \cite{McEwen2013Sparse, Wallis2017Sparse} for a more
comprehensive overview on this topic.  Somewhat restricted investigations of some aspects have already
been conducted, \textit{e.g.} considering sparsity in spherical harmonic space \cite{Rauhut2010},
sparsity in various redundant dictionaries \cite{Abrial2007, McEwen2013Sparse, Wallis2017Sparse}.

Variational inference techniques to solve inverse problems may be constructed in either the analysis or synthesis setting where
signal coefficients or coefficients of a sparse representation are recovered respectively \cite{Elad2006}.
For Euclidean settings the analysis problem typically provides greater reconstruction fidelity;
a characteristic often attributed to the lower cardinality of the analysis solution space \cite{Elad2006, Cleju2012choosing,
  Nam2013cosparse}, however comparisons between the analysis and synthesis settings on the sphere are not so clear, due to the approximate effective cardinality of different spaces on the sphere \cite{Wallis2017Sparse}.
There also exists a more fundamental binary-classification of optimization problems: constrained and
unconstrained, corresponding to regularization \textit{via} hard and soft constraints
respectively \cite{Boyd2004Convex}. Hard constraints (constrained formulation)
do not depend on variables such as Lagrangian multipliers, the optimal selection of which is an open problem, and instead constrain the solution to a certain sub-space. Soft constraints (unconstrained formulation)
can be considered as Bayesian inference problems \cite{Robert2001}
and thus support a principled statistical interpretation \cite{feeney:2014, Pereyra2017MAPUQ, Cai2018UQ}.

Traditionally, although variational approaches may support a probabilistic interpretation they
typically recover point estimates and do not quantify uncertainties. Fully probabilistic approaches (\emph{e.g.} Markov chain Monte Carlo sampling methods)
exist but are computationally expensive in the high dimensional setting of the sphere,
motivating the development of hybrid techniques. Recent developments in the field of
probability density theory \cite{Pereyra2017MAPUQ} address precisely this
consideration, facilitating flexible generation of scalable, fully principled Bayesian uncertainty
quantification (UQ) techniques for variational approaches. Many such techniques have been
developed \cite{Cai2018UQ, Repetti2018, M1, M2, M3}, with applications in a variety of
domains. In this article, we leverage these UQ techniques to recover
Bayesian local credible intervals, in effect pixel-level error bars, and other forms of hypothesis tests on discrete spherical
spaces. Interestingly we show how these uncertainties for a variety of common objectives can be computed rapidly (by exploiting
linearity) and in some cases analytically. Such computational savings are a key component
for the future of scalable UQ for spherical inverse problems.
Looking forward one might note that these UQ techniques for variational imaging rely only
on log-concavity of the posterior (convexity of the objective), as such a great many combinations of
likelihood (data-fidelity) and prior (regularization functionals) are permissible.

In the spirit of open access software and scientific reproducibility the spherical reconstruction
software (S2INV) developed during this project is made publicly available.\footnote{\url{https://github.com/astro-informatics/s2inv}}
S2INV is an object oriented C++ software package (with python extensions) which acts as a spherical extension to
the SOPT \cite{Onose2016SOPT, Carrillo2013SOPT, Carrillo2012SARA} software package
for flexible, efficient sparse optimization. We use fast exact spherical harmonic \cite{McEwen2011SSHT} and spherical
wavelet transforms \cite{Leistedt2013S2LET} to rapidly solve linear and ill-conditioned spherical inverse
problems. 

The remainder of this article is structured as follows. In Section \ref{sec:SphericalMaths} we
provide the mathematical context which underpins analysis of spin signals on the sphere.
In Section \ref{sec:inverse_problems} we present variational regularization approaches to solve 
spherical inverse problems and consider the unconstrained and constrained formulations, in both 
the analysis and synthesis settings. Furthermore, we discuss the generalization of planar regularization functionals
to their spherical counterparts, and briefly highlight highly optimized, scalable spherical reconstruction
open-source software available as a bi-product of this work.
In Section \ref{sec:UQ} we develop principled Bayesian uncertainty quantification  techniques
which can be leveraged for spherical inverse problems, and present acceleration methods exploiting function linearity and/or objective analytic solutions. A diverse selection
of numerical experiments are presented in Section \ref{sec:experiments} before providing
concluding remarks in Section
\ref{sec:conclusions}.


\section{Spin-signals on the sphere and rotation group} \label{sec:SphericalMaths}
One often wishes to consider the frequency space representation
of signals; whether this be embedded within regularization methods, necessary to fully capture a
desired forward model, or simply adopted to exploit computational symmetries (\emph{e.g.}
fast convolution algorithms). In the Euclidean setting, the frequency information of a signal is
efficiently expressed through projection onto Fourier space, the \emph{Fourier transform}.
For spherical settings frequency information is expressed though projection onto the space
of \emph{spin spherical harmonics}. In this section we review mathematical background 
fundamental to the analysis of signals defined on the sphere.

\subsection{Spin spherical harmonic transforms}
The space of square integrable spin-$s$ functions ${}_sf \in \mathsf{L}^2[\mathbb{S}^2]$,
for $s \in \mathbb{Z}$, with inner product $\langle \cdot | \cdot \rangle_{\mathbb{S}^2} $, are defined by their response
under local rotations of $\chi \in [0, 2\pi)$ about the tangent plane centered on the spherical
co-ordinate $\omega = (\theta, \psi) \in \mathbb{S}^2$, given by ${}_sf^{\prime}(\omega) = \text{e}^{-is\chi} {}_sf(\omega)$ where
${}_sf^{\prime}$ is the rotated function \cite{Newman1966SpinSignals, Goldberg1967Harmonics}. Such 
functions are most naturally represented by the
spin-weighted spherical harmonics ${}_sY_{\ell m} \in \mathsf{L}^2[\mathbb{S}^2]$ which
are a set of complete and orthogonal basis functions for degree $\ell \in \mathbb{Z}^+$
and integer $m \in \mathbb{Z}, \abs{m} \leq \ell,\abs{s} \leq \ell$.  We adopt
the Condon-Shortley phase convention \cite{condon1951theory}, which results in conjugate
symmetry ${}_sY^{\star}_{\ell m}(\omega) = (-1)^{s+m} {}_{-s}Y_{\ell -m} (\omega)$,
where $(\cdot)^{\star}$ denotes complex conjugation.
\par
A spin-$s$ function ${}_sf \in \mathsf{L}^2[\mathbb{S}^2]$ may be decomposed into the
spin spherical harmonic basis by
\begin{equation}
  {}_sf_{\ell m} \equiv \langle {}_sf_{\ell m} | {}_sY_{\ell m} \rangle_{\mathbb{S}^2} = \int_{\mathbb{S}^2} d\Omega(\omega) \; {}_sf(\omega) \; {}_sY_{\ell m}^{*}(\omega),
\end{equation}
\noindent
where $d\Omega(\omega) = \sin(\theta) d\theta d\psi$ is the standard rotation
invariant measure (Haar measure) on the sphere. Equivalently, by the orthogonality and completeness of ${}_sY_{\ell m}(\omega)$, 
one can exactly synthesize the signal space representation by
\begin{equation}
  {}_s f (\omega) = \sum^{\infty}_{\ell = 0} \sum^{\ell}_{m=-\ell} {}_sf_{\ell m} \; {}_sY_{\ell m}(\omega),
\end{equation}
\noindent
where the sum over $\ell$ is often truncated at $L$, where it is assumed that ${}_sf_{\ell m} = 0, \forall \ell \geq L$. In this sense, signals are considered to be bandlimited at $L$.
For notation brevity we adopt the shorthand operator notation $\bm{\mathsf{Y}}$ and $\bm{\mathsf{Y}}^{-1}$ to denote the forward and inverse spherical harmonic transforms.

This transformation allows one to probe the frequency content of spin signals
defined on the sphere, which facilitates, \emph{e.g}, efficient convolutions over spherical manifolds,
in much the same way one can compute convolutions over $\mathbb{R}^2$ through the
Fourier convolution theorem. In many cases signals have 0 spin, and so these relations
collapse to the simpler form most readers are likely familiar with. Nevertheless, a variety of interesting
physical settings exist where signals exhibit non-zero spin, \emph{e.g.} weak gravitational lensing, 
the cosmic microwave background, or quantum mechanical systems. 

\subsection{Scale-discretized directional spherical wavelets} \label{sec:wavelets}
Leveraging the above spin-$s$ spherical harmonic basis, and the associated convolutional
properties, one can construct wavelet dictionaries naturally on the sphere. To do
so one must first define a general rotation $\R_\rho$, for Euler angles
$\rho = (\alpha, \beta, \gamma) \in \text{SO}(3)$ with $\alpha \in [0, 2\pi), \beta \in [0, \pi)$,
and $\gamma \in [0, 2\pi)$, with action
$(\R_\rho {}_s f )(\omega) \equiv e^{-is\theta} {}_s f (\R_\rho^{-1} \bm{\omega})$. The
directional scale-discretized wavelet coefficients of any square integrable spin-$s$ function
${}_sf \in \mathsf{L}^2[\mathbb{S}^2]$ are given for scale $j$ by the directional convolution
\begin{align} \label{eq:directional_convolution}
  W^{{}_s\Psi^{(j)}} (\rho) & \equiv ( {}_sf \circledast {}_s\Psi^{(j)} ) (\rho) \equiv \langle {}_sf, \mathcal{R}_\rho {}_s\Psi^{(j)} \rangle_{\mathbb{S}^2} \nonumber \\
                            & = \int_{\mathbb{S}^2} d\Omega(\omega)
  {}_sf(\omega) (\mathcal{R}_\rho {}_s\Psi^{(j)})^\star (\omega),
\end{align}
where $\circledast$ represents the directional spherical convolution and ${}_s\Psi^{(j)} \in
  \mathsf{L}^2[\mathbb{S}^2]$ is the wavelet kernel at scale $j \in \mathbb{Z}^+$, which
determines the compact support of a given wavelet scale \cite{Leistedt2013S2LET}.
\par
Typically wavelet coefficients have negligible energy concentration over the low-frequency
domain in harmonic space, hence a scaling function ${}_s \Upsilon \in \mathsf{L}^2[\mathbb{S}^2]$
is introduced \cite{McEwen2015SpinWavelets, mcewen:s2let_ridgelets} with coefficients
$W^{{}_s\Upsilon} \in \mathsf{L}^2[\mathbb{S}^2]$ defined by the axisymmetric convolution
$\odot$ with a signal ${}_sf \in \mathsf{L}^2[\mathbb{S}^2]$ such that
\begin{align}
  W^{{}_s\Upsilon} (\omega) & \equiv ({}_sf \odot {}_s\Upsilon)(\omega) \equiv \langle {}_sf, \mathcal{R}_\omega {}_s\Upsilon \rangle_{\mathbb{S}^2} \nonumber \\
                            & = \int_{\mathbb{S}^2}
  d\Omega(\omega^\prime) {}_sf(\omega^\prime)(\mathcal{R}_\omega {}_s\Upsilon)^\star
  (\omega^\prime)\, ,
\end{align}
where $\R_\omega = \R_{(\psi, \theta, 0)}$ is a simplification of $\R_\rho$. One can straightforwardly show that 
the pixel-space representation of signals may be exactly synthesized 
if, and only if, the wavelet admissibility condition holds (see \emph{e.g.} \cite{McEwen2015SpinWavelets}).
There exist many functions which are admissible, \emph{e.g.} spherical
needlets \cite{marinucci:2008}, ridgelets \cite{mcewen:s2let_ridgelets} and curvelets \cite{Chan2017Curvelets}, however in this work
we choose adopt the directional scale-discretized wavelet harmonic space kernel \cite{Wiaux2008WaveletSynthesis,
  McEwenS2DWLocalisation}, For notational brevity we define operators
$\bm{\mathsf{\Psi}}$ and $\bm{\mathsf{\Psi}}^{-1}$ for the synthesis and analysis wavelet transforms respectively,
with corresponding adjoints operators $\bm{\mathsf{\Psi}}^{\dagger}$ and $(\bm{\mathsf{\Psi}}^{-1})^{\dagger}$ 
(for further details see, \emph{e.g.}, \cite{Wallis2017Sparse}).


\section{Generalized spherical imaging} \label{sec:inverse_problems}

Imaging inverse problems are found in countless areas of both science and industry; consequently
a great wealth of effort has been spent developing signal processing, Bayesian
inference and, more recently, machine learning techniques for solving such problems. However,
these techniques have overwhelmingly been restricted to Euclidean settings, in large part
due to their prevelance and relative simplicity.

As such, planar imaging benefits greatly from the flexibility such a dictionary of techniques affords,
whereas techniques developed for non-Euclidean manifolds (\emph{e.g.} the sphere) are
comparatively rare. One might reasonably consider applying planar techniques to
spherical settings, \emph{e.g.} through the analysis of planar projections, however these
fundamentally fall short \cite{Wallis2017MASSMAPPY} as a result of Gauss'
\emph{Theorema Egregium} --- a core concept of differential geometry, which dictates that
one may not flatten a ball without incurring significant distortions. Nonetheless, one can
certainly consider the development of analogous techniques defined natively on the sphere.
Previously, the spherical total variation TV-norm was constructed \cite{McEwen2013Sparse},
and the analysis and synthesis settings were compared in a spherical setting \cite{Wallis2017Sparse}.
In this section we extend the discussion to include the constrained and unconstrained formulations,
supported by a variety of proximal optimization algorithms, and a variety of regularization functionals.

On the sphere the setup of such imaging problems is as follows: consider the
case in which one acquires complex measurements $\bm{y} \in \mathbb{C}^M$, which may
or may not be natively on the sphere, but can be related to an estimated or true spherical signal
$\bm{x} \in \mathbb{C}^{N_{\mathbb{S}^2}}$ through the linear mapping
\begin{equation}
  \bm{\mathsf{\Phi}} \in \mathbb{C}^{M \times N_{\mathbb{S}^2} } : \bm{x} \in \mathbb{C}^{N_{\mathbb{S}^2}}
  \mapsto \bm{y} \in \mathbb{C}^M,
\end{equation}
commonly referred to as the measurement operator, which simulates measurement acquisition.
Suppose observations are contaminated with stochastic noise $\bm{n} \in \mathbb{C}^M$ such that $\bm{y} = \bm{\mathsf{\Phi}} \bm{x} + \bm{n}$
which is classically ill-posed. Furthermore,
when one considers that spherical observations are often incomplete, \emph{i.e.} $| M | \ll | N_{\mathbb{S}^2} |$,
such problem instances quickly become (seriously) ill-conditioned.
A diverse set of techniques exist to solve such inverse problems. This article is primarily concerned
with variational approaches, for which we develop uncertainty quantification techniques in Section \ref{sec:UQ}.
Variational approaches consider the inverse problem as a
minimization problem over a chosen objective function, which is typically the combination
of a data fidelity term and a regularization term --- selected to stabilize
reconstruction with \emph{a priori} assumptions as to the nature of the problem instance. Given
an objective function over which to minimize, one must make a variety of decisions regarding optimization formulation.

\subsection{Constrained and unconstrained optimization} \label{sec:optimisation_setting}
Suppose one selects data-fidelity term $f(\bm{x})$ and regularization functional $g(\bm{x})$, then
the unconstrained optimization problem has the Lagrangian formulation
\cite{Boyd2004Convex}
\begin{equation} \label{eq:unconstrained_optimisation}
  \bm{x}^\star  = \argminT_{\bm{x} \in \Omega} \big \lbrace f(\bm{x}) +
  \lambda g(\bm{x}) \big \rbrace,
\end{equation}
where $\Omega = \lbrace \mathbb{C}^{N_{\mathbb{S}^2}}$, $\mathbb{R}^{N_{\mathbb{S}^2}}$,
$\mathbb{R}^{N_{\mathbb{S}^2}}_+ \rbrace$, and the regularization parameter
$\lambda \in \mathbb{R}^+$ is a Lagrangian multiplier that balances the relative contributions
of the two functions to the objective. In effect $\lambda$ allows for a smooth re-weighting
(soft constraint) of the solution space instead of the strict boundary (hard constraint)
imposed in the constrained problem. When one formulates such optimizations in the unconstrained 
setting, the solution which minimizes the objective is
in fact the \emph{maximum a posteriori} (MAP) estimator $\bm{x}^\star = \bm{x}^{\text{MAP}}$.

Interestingly, it is well known that the unconstrained problem has direct links to Bayesian inference and supports
a principled statistical interpretation. However, until recently such
Bayesian interpretations have been restricted to point estimators and/or severely
restricted objective functional forms. One can leverage recent advances in probability concentration
theory \cite{Pereyra2017MAPUQ} to develop unconstrained optimization techniques which
support principled uncertainty quantification, as discussed in Section \ref{sec:UQ}.
Therefore, when considering spherical imaging problems, where Bayesian sampling
methods are impractical, in scientific domains, where uncertainty quantification is a desirable
feature, unconstrained optimization exhibits significant advantages.
However, this advantage comes at the additional complexity of optimal selection of the regularization parameter $\lambda$.
Popular methods for selection of $\lambda$ have adopted criteria such as: the Akaike information
criterion (AIC) \cite{akaike1998information}, Bayesian information criterion (BIC) \cite{schwarz1978estimating},
or Stein's unbiased risk estimator (SURE) \cite{stein1981estimation, pesquet2009sure} and others \cite{vidal:2018:maximum}. 
Optimal regularization parameter selection is still very much an open problem (for
various reasons including bias \textit{vs} variance considerations).
In this work we adopt a recently developed hierarchical Bayesian inference approach
\cite{Pereyra2015MMRegularisation} which treats the regularization parameter as a
nuisance variable \cite{Robert2001} over which a majorization-minimization algorithm
marginalizes. Effectively this method produces automatic, somewhat robust $\lambda$ selection
with a straightforward, natural Bayesian interpretation, facilitating principled uncertainty
quantification. 

Suppose instead that one is unwilling to accept a trade-off in either the data-fidelity or regularization functional, \emph{i.e.} one
requires that the data-fidelity is strictly below a given threshold, or that solutions belong to a restrictive sub-space of the regularization support or measurement operator.
For such inverse problems, the
problem instance is formulated as a constrained optimization problem, in which one function is minimized subject to the
constraint that the other function belongs to some constrained set \cite{Boyd2004Convex}. Here we consider
the common form in which the regularization functional is minimized subject to the constraint that the solution belongs within a level-set of the data-fidelity term, \emph{i.e.}
\begin{equation} \label{eq:constrained_optimisation}
  \bm{x}^\star = \argminT_{ \bm{x} \in \Omega} \big\lbrace g(\bm{x}) \big\rbrace
  \quad s.t. \quad f(\bm{x}) \leq \delta,
\end{equation}
where $\delta$ is a specified threshold (defining an iso-contour or level-set) of the data-fidelity term, typically
determined by the noise variance. This optimization restricts solutions to the sub-space $\bm{x} \in \mathcal{B}^\delta_{f}$
where $\mathcal{B}^\delta_{f}$ is the $f$-ball centered at $\bm{z} \in \Omega$ with radius $\delta$, \emph{i.e.} $\mathcal{B}^\delta_{f} (\bm{z}) \coloneqq \lbrace \bm{x}: f(\bm{x}) \leq \delta \rbrace.$

This formulation of the constrained problem requires calibration of $\delta$ which can be computed from the
estimated noise variance, and has a well defined interpretation. The calibration of
additional Lagrangian multipliers (regularization parameters) is not required, hence the
constrained setting is typically more straightforward to adopt. For many problem instances the constrained
setting provides greater reconstruction fidelity, though this is likely to be problem dependent.
In this sense the soft constraint adopted by the unconstrained setting (when
selected appropriately) allows for bias to be traded for variance (and \textit{vice versa})
and thus in particularly ill-posed problem instances, where the prior weighting is large (\textit{i.e.} high
bias situations), may produce estimates that are more accurate.
Furthermore, the constrained problem does not have an associated or well defined posterior
distribution over the latent space, prohibiting principled uncertainty quantification.

\subsection{Analysis and synthesis settings} \label{sec:analysis_synthesis}
Often one adopts regularization functions which are computed on projections of the image
space, \emph{e.g.} wavelet space, harmonic space, gradient space \emph{etc.}.
In such settings one can formulate optimization problems that consider the inverse problem
latent space to be the image space or the projected space, giving rise to the analysis and
synthesis formulations respectively. In this way one recovers solutions in pixel-space
$\bm{x}^\star$  (analysis) or projected space $\bm{\alpha}^\star$, which are then inverted to form pixel-space estimates
$\bm{x}^\star = \bm{\mathsf{\Psi}}\bm{\alpha}^\star$ (synthesis) \cite{Elad2006,
  Cleju2012choosing, Nam2013cosparse}.

This is most easily illustrated by considering a simple example. Consider the wavelet Lasso regression problem in the analysis form, \emph{i.e.}
an $\ell_1$ wavelet regularization functional
$g_{\text{A}}(\bm{x}) = \norm{\bm{\mathsf{\Psi}}^{-1} \bm{x}}_1$
and an $\ell_2$ data-fidelity term $f_{\text{A}}(\bm{x}) = \norm{\bm{\mathsf{\Phi}}\bm{x} - \bm{y}}_2^2$. Clearly in the analysis formulation the optimization problems are precisely
those given in Section \ref{sec:optimisation_setting}. However, in the synthesis settings the regularization functional takes the form $g_{\text{S}}(\bm{\alpha}) = \norm{\bm{\alpha}}_1$, while the data-fidelity term is given by $f_{\text{S}}(\bm{\alpha}) = \norm{\bm{\mathsf{\Phi}}\bm{\mathsf{\Psi}}\bm{\alpha} - \bm{y}}_2^2$.
With these definitions the synthesis optimization problem reads in much the same way as those presented in Section \ref{sec:optimisation_setting}
and, in fact, for situations in which the measurement operator is orthogonal, \textit{i.e.} $\bm{\mathsf{\Psi}}^{-1} = \bm{\mathsf{\Psi}}^{\dagger}$, these formulations are equivalent.  However, they have very different geometric properties when this is not the case \cite{Elad2006, Cleju2012choosing, Nam2013cosparse}.  
Notice that we adopt overcomplete spherical wavelet transforms where $\bm{\mathsf{\Psi}}^{-1} \not= \bm{\mathsf{\Psi}}^{\dagger}$, and sampled spherical harmonic transforms which are not orthogonal, \textit{i.e.} $\bm{\mathsf{Y}}^{-1} \not= \bm{\mathsf{Y}}^{\dagger}$ \cite{McEwen2013Sparse} --- a notable difference 
to the discrete Fourier transform in Euclidean settings. Therefore on the sphere the
analysis and synthesis settings are not equivalent, and often produce noticeably different results.

In practice the analysis setting has consistently
been demonstrated to exhibit greater reconstruction fidelity, a feature attributed to the
lower cardinality of the analysis solution space \cite{Elad2006, Cleju2012choosing, Nam2013cosparse}.
However, in previous work it was concluded that this characteristic may not generalized to the spherical setting \cite{Wallis2017Sparse}.
In Section \ref{sec:experiments} we revisit this analysis and find that the variation in relative
performance, both in terms of reconstruction fidelity and computational efficiency, of each
setting is dependent on the problem instance under consideration. Therefore,
flexibility with respect to reconstruction formulation supports development of
scalable spherical imaging algorithms tailored for specific applications. In this work we discover that implicit bandlimiting is often a determining factor when one considers inverse problems on the sphere, which impacts the effective cardinality of the spaces considered. In this sense it is beneficial to either (i) adopt the
synthesis setting in which signals are implicitly bandlimited during reconstruction or (ii) explicitly bandlimit the analysis setting, which is some settings can be computationally
inefficient on the sphere.
An example of such computational savings in the synthesis setting is demonstrated in Section \ref{sec:cmb_temp},
where an iterative Wiener filtering approach is adopted. In this case the analysis/synthesis formulations require $6,2$ 
spherical harmonic transforms respectively.

\subsection{Regularization functionals on the sphere} \label{sec:regs} \label{sec:regularisation_functions}
Having discussed the variety of ways one may formulate and construct an optimization on the
sphere, we should now consider spherical counterparts to common regularization functionals and
how one can develop these for the spherical setting. Such regularization functionals include
\emph{e.g.} sparsity promoting $\norm{\cdot}_1$ regularizers, typically in a sparsifying dictionary $\bm{\mathsf{\Psi}}$,
which are often motivated by the theory of compressed sensing \cite{Donoho2006,
  candes2006compressive, candes2006robust}; Gaussian $\norm{\cdot}_2^2$ regularizers,
which are often iterative implementations of harmonic Wiener filters \cite{hillery:1991, Kodi2019}; and
spherical total-variation (TV) priors \cite{McEwen2013Sparse}, which are effective
for edge detection and segmentation tasks.

Most imaging problems exist in the discrete settings, and so depend on approximations to
the underlying continuous $\ell_p$-norms. In spherical settings one often adopts equiangular
sampling \cite{McEwen2011SSHT}, which does not uniformly sample the continuous norms.
Typically this results in disproportionate weight being applied to pixels located at the poles, due
to progressing increased sampling density away from the equator. To account
for this spherical (directional wavelet) counter-parts to the traditional norms are defined by
\begin{equation}
  \snorm{\bm{x}}_p = \norm{\bm{w} \circ \bm{x}}_p \Rightarrow
  \pnorm{\bm{\alpha}}_p =  \big ( \sum_j \sum_n \snorm{\bm{\alpha}_{j, n} }_p^p \big )^{\frac{1}{p}},
\end{equation}
respectively, where $\bm{w} \in \mathbb{S}^2$ is the corresponding map of reciprocal pixel areas on the sphere, $\circ$ is the Hadamard product,
and $j, n \in \mathbb{Z}^+$ are wavelet scale and direction respectively. This reformulation provides a closer approximation to the underlying
continuous $\ell_p$-norm on the sphere.

With these corrected norms one can straightforwardly consider, \emph{e.g.}, sparsity in spherical wavelet space $\pnorm{\bm{\mathsf{\Psi}}^\dagger \bm{x}}_1$. Such a generalization
permits multi-resolution algorithms \cite{Leistedt2013S2LET} resulting in wavelet scale
projections of varying resolution, which provide a significant increase in computational
efficiency, a fundamental bottleneck of variational methods on the sphere.
In theory one could leverage the exact quadrature weights inherent to the underlying spherical sampling theorems \cite{McEwen2011SSHT}, however in this work we find simple weights $\bm{w}_{j, n}$ which capture the pixel area to be sufficient.

\subsection{Efficient flexible imaging on the sphere} \label{sec:efficient_algorithms}
Variational approaches efficiently locate optimal solutions \emph{via} iterative algorithms, which typically leverage $1^{\text{st}}$-order (gradient) information to navigate towards extremal values. Furthermore, for convex objectives, such algorithms permit strong guarantees of both convergence and the rate of convergence. Imaging problems often adopt non-differentiable regularization functionals (\emph{e.g.} $\ell_1$-norms) for which proximal operators may be used to navigate the objective function, thus motivating proximal convex optimization algorithms.

Convex optimization algorithms require successive iterations
to converge; as such, any operators evaluated must be
efficient and precise, so as to facilitate accurate, scalable methods. These considerations
are more pronounced when considering optimization over spherical manifolds, wherein
underlying operators (\emph{e.g.} spin-$s$ spherical harmonic transforms) scale poorly
with dimension ($\propto \mathcal{O}(L^3)$ in the best case scenario).
Additionally, a large subset of optimization algorithms require adjoint $\dagger$ operators,
which are often incorrectly approximated by their inverse operators,
introducing unpredictable errors and breaking convergence guarantees.
Furthermore, on the sphere one must also consider the weighting scheme presented in Section~\ref{sec:regs}, 
which can be incorporated into proximal optimization algorithms through; a direct operator that performs the weighting, or by weighting norms.
To avoid additional complications (\emph{e.g.} under certain norms weighting operators do not represent tight frames, necessitating additional 
sub-iterations) we simply weight the norms directly.

During this research we developed a highly optimized object oriented (OOP) C++ software framework (S2INV) which permits all the aforementioned flexibility. The
equiangular sampling theorem on the sphere of \cite{McEwen2011SSHT} is adopted through the SSHT\footnote{\url{https://astro-informatics.github.io/ssht/}} package, which permits fast and efficient spin-$s$ spherical harmonic transforms, whilst permitting
machine precision computation. Additionally, we adopt optimized scale-discretized directional
wavelets on the sphere \cite{Wiaux2008WaveletSynthesis,mcewen:2013:waveletsxv,
  McEwen2015SpinWavelets} through the S2LET\footnote{\url{https://astro-informatics.github.io/s2let/}} package \cite{Leistedt2013S2LET, Wallis2017Sparse}, which
are optimally sampled and support machine precision synthesis. We leverage a
recently developed, highly optimized C++ OOP sparse optimization framework SOPT\footnote{\url{http://astro-informatics.github.io/sopt/}}
\cite{Onose2016SOPT, Carrillo2013SOPT, Carrillo2012SARA}, which facilitates a variety of
proximal convex optimization algorithms, \emph{e.g.} forward-backward
\cite{Combettes2011Proximal,beck2009fast}, primal dual
\cite{Boyd2004Convex,Komodakis2015, Combettes2014}, and the alternating direction method of multipliers \cite{boyd2011distributed}, with appropriate modifications for the spherical setting.
In this way S2INV provides a scalable, flexible, open-source software package, which is fully customizable
and supports a wide variety of novel, fully principled, uncertainty quantification techniques on the sphere,
which we discuss in Section \ref{sec:UQ}.


\section{Spherical Bayesian uncertainty quantification} \label{sec:UQ}

The unconstrained reconstruction problem has a straightforward Bayesian interpretation which is as follows.
The posterior distribution of a spherical image $\bm{x} \in \mathbb{C}^{N_{\mathbb{S}^2}}$
defined over, \emph{e.g.}, the celestial sphere or the globe, given observations
$\bm{y} \in \mathbb{C}^M$ is given by Bayes' theorem,
\begin{equation} \label{eq:bayes_theorem}
  P(\bm{x}|\bm{y}; \mathcal{M}) \equiv \frac{P(\bm{y}|\bm{x}; \mathcal{M})
  P(\bm{x}; \mathcal{M})}{ \int_{\mathbb{C}^{N_{\mathbb{S}^2}}}
  P(\bm{y}|\bm{x}; \mathcal{M})P(\bm{x}; \mathcal{M})d\bm{x} },
\end{equation}
where the likelihood encodes data fidelity, the prior encodes \textit{a priori} information of the image, and
$\mathcal{M}$ represents some model, which includes the mapping
$\bm{\mathsf{\Phi}} \in \mathbb{C}^{M \times N_{\mathbb{S}^2}} : \bm{x} \mapsto \bm{y}$,
and some understanding of the noise inherent to $\bm{y}$ \cite{Robert2001}.
Note that the marginal likelihood (Bayesian evidence) is a constant scaling of the posterior
and can be used for model comparison, which we do not consider further in this article.

Typically sampling methods, \emph{e.g.} Markov chain Monte Carlo, are adopted to sample from the posterior distribution from which one can determine a point estimation of the solution to the inverse problem and the
distribution of uncertainty about such a solution. Although these methods recover asymptotically exact
estimates of the posterior distribution, they typically require large numbers of samples to converge. Each sample requires at least a single evaluation of the posterior
which in spherical settings is computationally demanding --- for
moderate resolutions $L > 10^3$ sampling methods rapidly become computationally
intractable.

Instead consider a variational approach that maximizes the posterior odds, referred to as the \textit{maximum a posteriori} (MAP) solution defined by
\begin{align}
  \bm{x}^{\text{MAP}} & \equiv \argmaxT_{\bm{x}} \big \lbrace P(\bm{x}|\bm{y};\mathcal{M})
  \big \rbrace, \nonumber                                                                                                    \\
                      & \propto \argminT_{\bm{x}} \big \lbrace -\log ( \; P(\bm{y}|\bm{x};\mathcal{M})P(\bm{x}; \mathcal{M})
  \;) \big \rbrace, \nonumber                                                                                                \\
                      & \propto \argminT_{\bm{x}} \big \lbrace  h(\bm{x}) = f(\bm{x}) + g(\bm{x}) \big \rbrace,
\end{align}
where the second line follows by the monotonicity of the
logarithm function. For convex objective functions $h(\bm{x})$ this takes the form of a convex optimization
problem \cite{Boyd2004Convex}, and therefore Equation \ref{eq:unconstrained_optimisation}
explicitly returns the MAP solution, as asserted in Section \ref{sec:optimisation_setting}.
Hence, leveraging state-of-the-art convex optimization techniques one can efficiently locate
the solution which maximizes the posterior odds. However this is still a point estimate which,
though useful, does not naively support uncertainty quantification. Recently, approximate contours of the 
latent space have been derived facilitating variational regularization methods with principled uncertainty quantification. 
We discuss these approximate methods and develop uncertainty quantification techniques on the sphere, which we accelerate by exploiting function linearity.

\subsection{Highest posterior density credible regions} \label{sec:HPD}

A credible region $C_{\alpha} \subset \mathbb{C}^{N_{\mathbb{S}^2}}$ of the posterior latent space
at credible confidence $100(1-\alpha)\%$, for $\alpha \in [0,1]$, satisfies the integral equation \cite{Robert2001}
\begin{equation} \label{eq:credible_region_integral}
  P(\bm{x} \in C_{\alpha}|\bm{y}; \mathcal{M}) = \int_{\bm{x} \in \mathbb{C}^{N_{\mathbb{S}^2}}} P(\bm{x}|
  \bm{y}; \mathcal{M})\mathbb{I}_{C_{\alpha}}d\bm{x} = 1 - \alpha,
\end{equation}
where $\mathbb{I}_{C_{\alpha}}$ is the standard set indicator function. The optimal credible region in the sense of
minimal volume \cite{Robert2001} is the highest posterior density (HPD)
credible region defined by $C_{\alpha} \coloneqq \lbrace \bm{x} : h(\bm{x}) \leq \epsilon_{\alpha} \rbrace$,
where $\epsilon_{\alpha} \in \mathbb{R}^+$ is an isocontour of the log-posterior. Determination of the HPD region requires computation of the integral in Equation \ref{eq:credible_region_integral}, which is
computationally infeasible in even moderate dimensional spherical settings, due to dimensionality and functional complexity considerations.
Convex objectives $h(\bm{x})$ support the conservative approximate HPD credible region
$C^{\prime}_{\alpha}$ defined by \cite{Pereyra2017MAPUQ}
\begin{align}
  C_{\alpha}  \subseteq C^{\prime}_{\alpha} \subset \mathbb{C}^{N_{\mathbb{S}^2}} & \coloneqq
  \Big \lbrace \bm{x} : h(\bm{x}) \leq \epsilon^{\prime}_{\alpha} \Big \rbrace, \nonumber                                                              \\
  \text{where} \quad \epsilon^{\prime}_{\alpha} = h(\bm{x}^{\text{MAP}})          & + \sqrt{16 N \log(3 / \alpha)} + N, \label{eq:approx_credible_set}
\end{align}
which allows one to approximate $C_\alpha$ with only knowledge of the MAP solution
$\bm{x}^{\text{MAP}}$ and the dimension $N = \mathbb{C}^{N_{\mathbb{S}^2}}$. An upper
bound on the approximation error exists \cite{Pereyra2017MAPUQ}. 
Therefore for convex objectives, given $\bm{x}^{\text{MAP}}$, one may draw statistically
principled conclusions. This credible set approximation has been leveraged to develop fast Bayesian uncertainty quantification techniques in a variety of settings \cite{Cai2018UQ, M1, M2, M3, Repetti2018, Pereyra2017MAPUQ}.

\subsection{Bayesian hypothesis testing on the sphere} \label{sec:hypothesis}
The most straightforward uncertainty quantification technique one may generate by leveraging the approximation of Equation \ref{eq:approx_credible_set} is that of hypothesis testing \cite{Cai2018UQ, M1, M4}. The concept of hypothesis testing is to adjust a feature of the recovered estimator $\bm{x}^{\text{MAP}}$ generating a surrogate solution $\bm{x}^{\text{sur}}$, of which we ask $\bm{x}^{\text{sur}} \in C^{\prime}_{\alpha}$?
If $\bm{x}^{\text{sur}} \not\in C^{\prime}_{\alpha} \Rightarrow \bm{x}^{\text{sur}} \not\in C^{\prime}$, which follows from the conservative nature of the approximation in Equation \ref{eq:approx_credible_set}, the feature of interest is considered to be statistically significant (necessary to the reconstruction) at $100(1-\alpha)\%$ confidence. Conversely $\bm{x}^{\text{sur}} \in C^{\prime}_{\alpha}$ indicates that the surrogate solution remains within the approximate credible set and we conclude that the feature is indeterminate.


\subsection{Local credible intervals on the sphere} \label{sec:LCI}
%
Suppose one recovers an optimal solution $\bm{x}^{\text{MAP}}$ through unconstrained convex
optimization (see Section \ref{sec:optimisation_setting}) and wishes to quantify the uncertainty
associated with a given pixel or super pixel (collection of pixels). With knowledge of the
approximate level set threshold $\epsilon^{\prime}_{\alpha}$, and therefore the approximate
HPD credible set $C^{\prime}_{\alpha}$ at well defined confidence $(100 - \alpha)\%$, one simply
needs to iteratively compute the extremal values a given region of interest may take,
such that the resulting solution falls outside of the approximate HPD credible set, \emph{i.e.} $\bm{x}^{\text{sur}} \not\in C^{\prime}_{\alpha}$. 
One must then define which types of regions (super-pixels) on the sphere one is interested in.

Formally, select independent partitions of the latent space $\Omega = \cup_i \Omega_i$
for which we define super-pixel indexing functions $\zeta_{(\cdot)}$ such that $\bm{x}_i \in \Omega_i \Rightarrow \zeta_{\Omega_i} = 1$ and $\bm{x}_i \notin \Omega_i \Rightarrow \zeta_{\Omega_i} = 0$.
For a given $\Omega_i$
locate upper (lower) bounds $\xi^+_{\Omega_i}$,  $\xi^-_{\Omega_i}$ respectively,
which saturate the HPD credible region $C^{\prime}_{\alpha}$ \cite{Cai2018UQ, M2}. This is 
achieved by the following optimizations,
\begin{align}
  \xi^{\pm}_{\Omega_i} & = \maxT_{\xi \in \mathbb{R}^+} \big \lbrace \pm \xi | f(\bm{x}_{i,\xi}) + g(\bm{x}_{i,\xi}) \leq \epsilon^{\prime}_{\alpha} \big \rbrace, \label{eq:lci_opt}
\end{align}
where $\bm{x}_{i,\xi} = \bm{x}^{\text{MAP}} \zeta_{\Omega /\Omega_i} + \xi \zeta_{\Omega_i}$
is a surrogate solution where the super pixel region has been replaced by a uniform intensity $\xi$.
The collective set of these bounds $\lbrace \abs{ \xi^+_{\Omega_i} - \xi^-_{\Omega_i} } \rbrace$
is taken to be the local credible interval map \cite{Cai2018UQ}, which can simply be recovered
\textit{via} bisection. Though conditional local uncertainty quantification techniques such as this have demonstrated utility in certain
circumstances \cite{Cai2018UQ, M1, M2, M3}, in the high dimensional spherical settings they can
quickly become dilute \cite{M4}. This makes intuitive sense, as small (local) objects (super-pixels)
in high dimensional settings become statistically insignificant. As such, in high-dimensional
spherical settings global or aggregate (statistical) uncertainty quantification techniques are
often more meaningful (see \emph{e.g} \cite{M4}).


\subsubsection{Gridding schemes}
One can construct rectangular partitions directly on the latent space (\emph{e.g.} uniform gridding). 
However, in the spherical setting it is sometimes more meaningful to define a super pixel by a 
fixed physical area surrounding a defined central
pixel. Practically this is computed as follows: define a central pixel on the sphere, rotate this pixel 
to a pole (where higher angular resolution provides greater
fine tuning of super-pixel area), select a given angular deviation from the pole, define this
spherical cap as the super pixel. In this way all super pixels are, by definition, of
equal area.

\subsection{Acceleration through linearity} \label{sec:linearity}
Naive computation of local credible intervals through bisection can often require many evaluations of the objective function, which is particularly costly when one considers functions on
the sphere. To avoid this computational bottleneck we exploit the linearity of such operators.
Consider the generalized convex objective function for the analysis setting, without loss of generality, which can be written as
\begin{equation} \label{eq:problem}
  h(\bm{x}) = f(\bm{x}) + g(\bm{x}) = \norm{\bm{\mathsf{\Phi}}\bm{x} - \bm{y}}_{p_1}^{p_2} + \norm{\bm{\mathsf{\Psi}}^\dagger \bm{x}}_{q_1}^{q_2}.
\end{equation}
Consider again the partition $\bm{x}^{\text{sur}} = \bm{x} \zeta_{\Omega/\Omega_i}
  + \xi \zeta_{\Omega_i}$ upon which the applications of any linear operator $\mathcal{L}$ is given
trivially by linearity to be
$\bm{\mathcal{L}}\bm{x}^{\text{sur}} = \bm{\mathcal{L}}\bm{x} \mathbb{I}_{\zeta_{\Omega/\Omega_i}} + \xi \bm{\mathcal{L}}\mathbb{I}_{\zeta_{\Omega_i}}$.
Explicitly expanding Equation \ref{eq:problem} with linearity one finds
\begin{align}
  \norm{\bm{\mathsf{\Phi}}\bm{x}\mathbb{I}_{\zeta_{\Omega/\Omega_i}} + \xi \bm{\mathsf{\Phi}}\mathbb{I}_{\zeta_{\Omega_i}} - \bm{y}}^{p_2}_{p_1}
                                                     & + \norm{\bm{\mathsf{\Psi}}^\dagger \bm{x}\mathbb{I}_{\zeta_{\Omega/\Omega_i}} + \xi \bm{\mathsf{\Psi}}^\dagger \mathbb{I}_{\zeta_{\Omega_i}} }^{q_2}_{q_1}, \nonumber \\
  \Rightarrow \norm{\bm{a} + \xi \bm{b}}^{p_2}_{p_1} & + \norm{\bm{c} + \xi \bm{d} }^{q_2}_{q_1},
\end{align}
for constant (per credible interval) vectors defined to be
\begin{align}
  \bm{a} & = \bm{\mathsf{\Phi}}\bm{x}\mathbb{I}_{\zeta_{\Omega/\Omega_i}} - \bm{y} & \quad
  \bm{b} & = \bm{\mathsf{\Phi}}\mathbb{I}_{\zeta_{\Omega_i}} \nonumber                     \\
  \bm{c} & = \bm{\mathsf{\Psi}}^\dagger \bm{x}\mathbb{I}_{\zeta_{\Omega/\Omega_i}} & \quad
  \bm{d} & = \bm{\mathsf{\Psi}}^\dagger \mathbb{I}_{\zeta_{\Omega_i}}
\end{align}
In this way the local credible optimization problems in Equation \ref{eq:lci_opt} can be
re-written instead as
\begin{equation} \label{eq:general_shape}
  \norm{\bm{a} + \xi \bm{b}}^{p_2}_{p_1} + \norm{\bm{c} + \xi \bm{d} }^{q_2}_{q_1}
  \leq \epsilon^{\prime}_{\alpha},
\end{equation}
which is clearly just a 1-dimensional polynomial root finding problem. One could approach
this inequality from an iterative perspective, forming an upper bound through the Minkowski 
inequality, which is then leveraged as initialization for bisection. For polynomials of order 
$< 5$ (see Abel-Ruffini theorem) Equation \ref{eq:general_shape} permits analytic
solutions. In practice the computational difference between the analytic solution and solving
an inequality bounded bisection problem is marginal, though in high dimensions
this speed up is non-negligible.

\subsubsection{Gaussian regression}
Suppose one adopts both a Gaussian likelihood and prior (\emph{e.g.} iterative Wiener
filtering approaches), in such a setting we have $p_1 = p_2 = q_1 = q_2 =2$ which reduces Equation 
\ref{eq:general_shape} to the binomial inequality
  $\norm{\bm{a} + \xi \bm{b}}^{2}_{2} + \norm{\bm{c} + \xi \bm{d} }^{2}_{2} \leq \epsilon^{\prime}_{\alpha}$,
which expands to give
\begin{equation}
  \big[ \norm{\bm{b}}_2^2 + \norm{\bm{d}}_2^2 \big] \xi^2
  + 2\big[ \bm{a} \cdot \bm{b} + \bm{c} \cdot \bm{d} \big] \xi
  + \big[ \norm{\bm{a}}_2^2 + \norm{\bm{c}}_2^2 \big]
  \leq \epsilon^{\prime}_{\alpha}.
\end{equation}
One could gain some geometric insight by considering the case in which $\bm{a} \cdot \bm{b} + \bm{c} \cdot \bm{d}=0$,
as in such a case the resulting credible region about the posterior is symmetric, however
we do not consider this further here. Nonetheless, per credible region the interval is trivially recovered.

\subsubsection{Lasso regression}
Consider the Lagrangian dual of the Lasso regression
(\emph{e.g.} sparse reconstruction), in which we have $p_1 = p_2 = 2$ and $q_1 = q_2 = 1$, such that
the general
polynomial Equation \ref{eq:general_shape} reduces to a $2^{\text{nd}}$-order polynomial
  $\norm{\bm{a} + \xi \bm{b}}^{2}_{2} + \norm{\bm{c} + \xi \bm{d} }_1 \leq \epsilon^{\prime}_{\alpha}$,
which, assuming the intersection of the partitions projected into $\bm{\mathsf{\Psi}}$ is negligible, \emph{i.e.}
the dictionary $\bm{\mathsf{\Psi}}$ has sufficient localization properties on the sphere, results in the inequality
\begin{equation}
  \norm{\bm{b}}_2^2 \xi^2
  + 2\big[ \bm{a} \cdot \bm{b} + \norm{d}_1 \big] \xi
  + \big[ \norm{\bm{a}}_2^2 + \norm{\bm{c}}_1 \big]
  \leq \epsilon^{\prime}_{\alpha},
\end{equation}
which can be analytically solved. Typically the partitions $\Omega_i$ projected into $\bm{\mathsf{\Psi}}$
exhibit overlapping support and so this inequality is not exact. In such cases one can
perform bisection, computing only the $\ell_1$ term at each iteration.

\section{Numerical experiments} \label{sec:experiments}
In this section we showcase the variational regularization and uncertainty quantification techniques presented in Section \ref{sec:inverse_problems} and Section \ref{sec:UQ} on a diverse set of numerical experiments.
For each scenario we create mock observations $\bm{y} = \bm{\mathsf{\Phi}}\bm{x}$ of a ground truth
signal $\bm{x}$ which are related through the forward model $\bm{\mathsf{\Phi}} : \bm{x} \mapsto \bm{y}$
from which we formulate an ill-posed inverse problem (\textit{e.g.} add noise, mask, blur, \textit{etc.}).
We solve this inverse problem for an estimator $\bm{x}^\star$ ($\bm{x}^{\text{MAP}}$) of
$\bm{x}$ the success of which we quantify by the recovered \textit{signal to noise ratio},
defined by
$\text{SNR} = 20 \log_{10} ({\norm{\bm{x}}_2}/{\norm{\bm{x} - \bm{x}^\star}_2} )$.

\subsection{Earth satellite topography} \label{sec:earth_topo}
Suppose a satellite performs observations $\bm{y} \in \mathbb{R}^{M}$ of the
Earth's topography (geographic elevation) which can be related to the true topography $\bm{x} \in \mathbb{R}^{N_{
      \mathbb{S}^2}}$ through a mapping (forward model) $\bm{\mathsf{\Phi}}_{\text{ET}} \in \mathbb{R}^{M \times N_{
      \mathbb{S}^2}} : \bm{x} \mapsto \bm{y}$.
Consider the scenario in which incomplete, \emph{i.e.} $\abs{M} \ll \abs{N_{\mathbb{S}^2}}$, observations $\bm{y}$ are contaminated with independent and identically
distributed (\textit{i.i.d.}) Gaussian noise $\bm{n} \sim \mathcal{N}(0, \sigma^2) \in \mathbb{R}^{M}$, and blurred with an axisymmetric smoothing kernel with full width half maximum (FWHM) $\Theta$. In such a case
observations are modeled by $\bm{y} = \bm{\mathsf{\Phi}}_{\text{ET}} \bm{x} + \bm{n}$ for measurement operator
\begin{equation}
  \bm{\mathsf{\Phi}}_{\text{ET}} = \bm{\mathsf{D}} \bm{\mathsf{Y}^{-1}} \bm{\mathsf{\Theta}} \bm{\mathsf{Y}}
  \quad \text{and} \quad \bm{\mathsf{\Phi}}_{\text{ET}}^\dagger = \bm{\mathsf{Y}}^\dagger
  \bm{\mathsf{\Theta}} \ (\bm{\mathsf{Y}}^{-1})^\dagger \bm{\mathsf{D}}^\dagger,
\end{equation}
where $\bm{\mathsf{Y}}, \bm{\mathsf{Y}}^{-1}$
are forward and inverse spherical harmonic transforms
correspondingly, $\bm{\mathsf{D}}, \bm{\mathsf{D}}^\dagger$ are masking
and projection operators correspondingly, $\bm{\mathsf{\Theta}}$ is the axisymmetric convolution with
the harmonic representation of $\Theta$ which is trivially self-adjoint, and $\dagger$ represents
the adjoint operation.

As $\bm{n}$ is a univariate Gaussian the data-fidelity (log-likelihood) term is simply given
by \mbox{$\frac{1}{2\sigma^2}\snorm{\bm{\mathsf{\Phi}}_{\text{ET}}\bm{x} - \bm{y}}_2^2$}.
Here we adopt a sparsity promoting $\ell_1$-norm wavelet regularization $\pnorm{\bm{\mathsf{\Psi}}^\dagger(\cdot)}_1$
(Laplace distribution log-prior), and solve both the constrained formulation through the proximal
ADMM algorithm \cite{boyd2011distributed} and the unconstrained formulation through the
proximal forward backward algorithm \cite{beck2009fast, Combettes2011Proximal, Boyd2004Convex},
in both the analysis and synthesis settings. Notice the use of spherical (wavelet) space norms,
outlined in Section \ref{sec:regularisation_functions}, which better approximate spherical continuous norms.
\begin{figure}
  \includegraphics[width=\columnwidth, trim={3cm 5.5cm 3cm 6cm},clip]{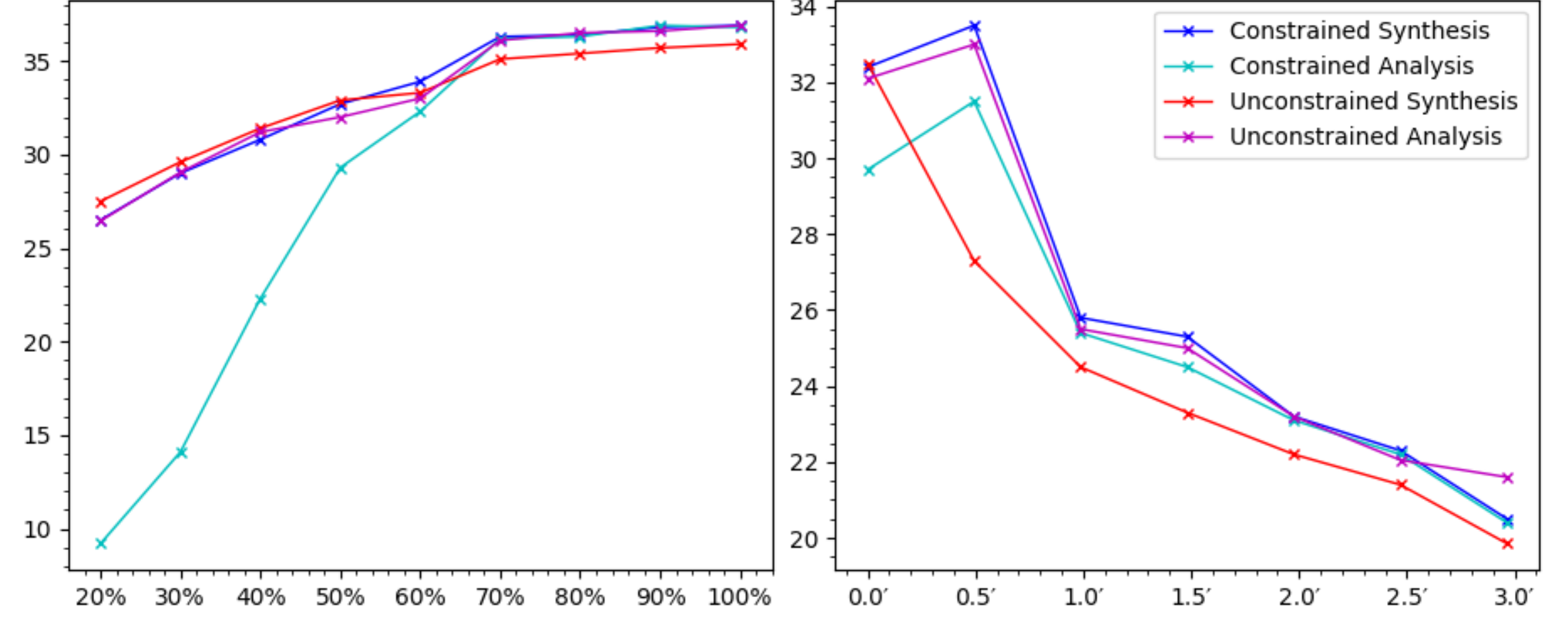}
  \put(-110,-8){\footnotesize Axisymmetric kernel scale $\Theta^{\prime}$}
  \put(-232,-8){\footnotesize Fraction of sphere observed}
  \put(-258,48){\rotatebox[origin=c]{90}{\footnotesize SNR (dB)}}

  \caption{
    \textbf{Left:} Recovered SNR for a variety of problem setups over a
    variety of inpainting scenarios versus \% of pixels observed.
    \textbf{Right:} Recovered SNR for a variety of problem setups
    versus variety of axisymmetric blurring kernel scales, for a fixed inpainting of 50\% masked
    pixels.
    \textbf{Discussion:} Generally each setup performs similarly and it would appear that
    no single setting is optimal in all cases. Notice the drastic underperformance of the analysis
    constrained formulation in the heavily masked regime for the inpainting problem. We find
    that this asymmetry is due to the implicit spherical harmonic bandlimiting inherent to the synthesis 
    problem (see text).
  }
  \label{fig:earth_topo_spread}
\end{figure}
To quantify the impact of analysis versus synthesis (and constrained versus unconstrained)
settings we consider all settings in two paradigms (i) varying levels of inpainting without
deconvolution (ii) varying scales of deconvolution with 50\% masked pixel inpainting. Generally, each problem setup performs comparably in all settings considered (see Figure \ref{fig:earth_topo_spread}). Certainly it
cannot be said that one reconstruction paradigm is optimal in all settings, which leads us
to conclude that it is likely that problem formulation optimality is ambiguous and should be selected on
a case by case basis. Interestingly, notice the underperformance of the constrained
analysis problem in the heavily masked regime (see the left plot of Figure
\ref{fig:earth_topo_spread}). This was observed in prior analysis \cite{Wallis2017Sparse} and reported as evidence that the synthesis
setting may produce more optimal results.

\begin{figure}
  \centering
  \begin{tikzpicture}[spy using outlines={circle, magnification=3, connect spies}]
    \node[anchor=south west,inner sep=0] (image) at (0,2.7) {
      \begin{minipage}{.49\columnwidth}
        \includegraphics[width=\columnwidth, trim={10cm 11cm 10cm 5.5cm},clip]{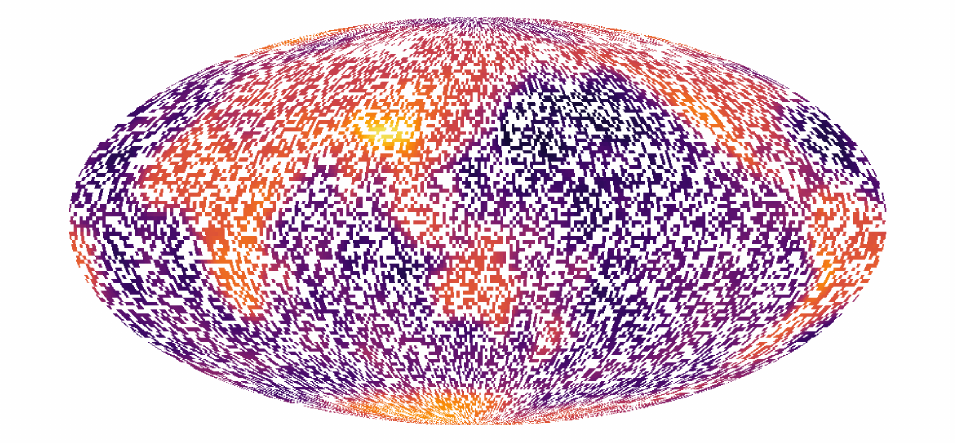}
        \put(-85, 62){Observations}
      \end{minipage}
      \begin{minipage}{.49\columnwidth}
        \includegraphics[width=\columnwidth, trim={10cm 5.5cm 10cm 11cm},clip]{topographic_example.pdf}
        \put(-90, 62){Reconstruction}
      \end{minipage}
    };
    \node[anchor=south west,inner sep=0] (image) at (0,0) {
      \hspace{.25\columnwidth}
      \begin{minipage}{.49\columnwidth}
        \includegraphics[width=\columnwidth, trim={9cm 2cm 9cm 12.2cm},clip]{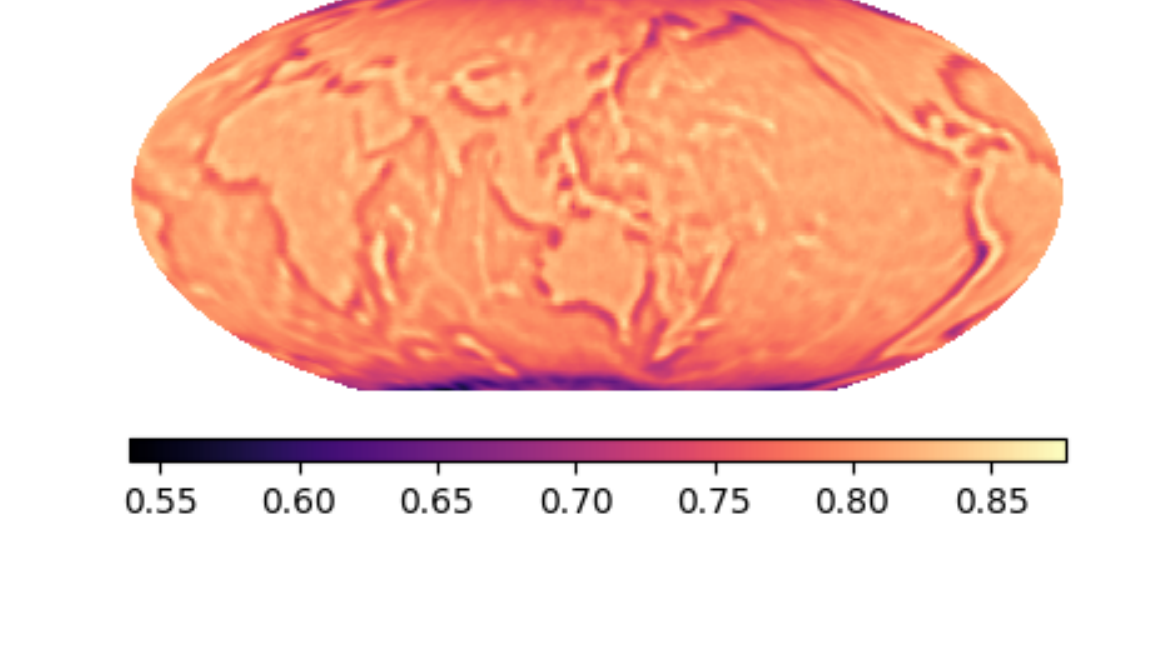}
        \put(-85, 75){Uncertainty}
      \end{minipage}
    };

    \spy [black, size=2.0cm] on (7.0,3.2)
    in node[fill=white] at (7.8,1.4);

    \spy [black, size=2.0cm] on (2.5,3.24)
    in node[fill=white] at (1.2,1.4);

  \end{tikzpicture}%
  \vspace{-25pt}
  \caption{\textbf{Left:} Simulated observations contaminated with 30dB Gaussian \textit{i.i.d.} noise,
    convolved with a $\sim 268$ arc-minute Gaussian blurring kernel, and with 60\% of
    pixels masked. \textbf{Right:} Unconstrained reconstruction using $\ell_1$-norm wavelet sparsity
    regularization (log-prior) in the analysis setting,
    adopting the proximal forward-backward algorithm \cite{beck2009fast, Boyd2004Convex, Combettes2011Proximal}.
    \textbf{Bottom:} Bayesian local credible intervals at $99\%$ confidence (reconstruction values are $\in [-1,1]$), which are pixel-level uncertainties (see Section \ref{sec:UQ}).}
  \label{fig:earth_topo}
  \vspace{3pt}
  \begin{tikzpicture}[spy using outlines={circle, magnification=4, connect spies}]
    \node[anchor=south west,inner sep=0] (image) at (0,0) {
      \begin{minipage}{.49\columnwidth}
        \includegraphics[width=\columnwidth, trim={10.6cm 11.1cm 10.6cm 5.7cm},clip]{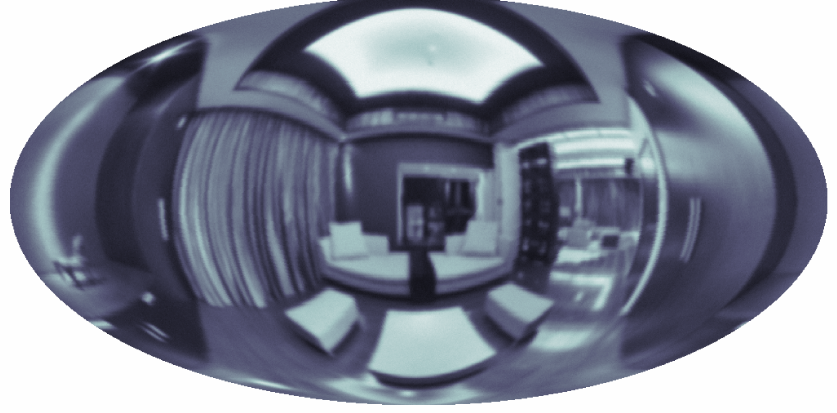}
        \put(-85, 65){Observations}
      \end{minipage}
      \begin{minipage}{.49\columnwidth}
        \includegraphics[width=\columnwidth, trim={10.6cm 5.7cm 10.6cm 11.1cm},clip]{camera_example.pdf}
        \put(-90, 65){Reconstruction}
      \end{minipage}
    };

    \spy [black, size=2.0cm] on (5.9,0.60)
    in node[fill=white] at (7.7,0.1);

    \spy [black, size=2.0cm] on (1.45,0.64)
    in node[fill=white] at (3.2,0.1);

  \end{tikzpicture}%
  \vspace{-23pt}
  \caption{\textbf{Left:} Simulated observations contaminated with 30dB Gaussian noise and
    convolved with a $\sim 78$ arc-minute Gaussian blurring kernel (\textit{e.g.} motion blur).
    \textbf{Right:} Spherical TV-norm $\norm{\bm{x}}_{\text{TV}}$
    regularized reconstruction through proximal primal dual algorithm in the analysis setting.}
  \label{fig:4pi_camera}

\end{figure}

Note that the synthesis setting implicitly bandlimits the observations $\bm{x}$,
therefore restricting the solution space cardinality --- a factor known to impact reconstruction
fidelity \cite{Elad2006}. To account for this bias we reran the analysis optimization with an explicitly 
bandlimited measurement operator, results of which can be seen in Figure \ref{fig:earth_topo} 
(which also demonstrates the uncertainty quantification of Section \ref{sec:LCI}). It was
found that the analysis setting performs similarly to the synthesis setting for this problem,
leading us to conclude that the optimality of optimization formulation is at best ambiguous.

\subsection{$360^{\circ}$ camera blur deconvolution}

Suppose a $360^{\circ}$ camera captures a greyscale spherical image $\bm{y} \in \mathbb{R}^{M_{
      \mathbb{S}^2}}$ which can be related to the true image $\bm{x} \in \mathbb{R}^{N_{
      \mathbb{S}^2}}$ through the forward model $\bm{\mathsf{\Phi}}_{360^{\circ}} \in \mathbb{R}^{M_{
      \mathbb{S}^2} \times N_{\mathbb{S}^2}} : \bm{x} \mapsto \bm{y}$.
Consider that the camera captures complete $\abs{M_{\mathbb{S}^2}} = \abs{N_{\mathbb{S}^2}}$ observations but introduces
low-level \textit{i.i.d.} Gaussian noise $\bm{n} \sim \mathcal{N}(0, \sigma^2) \in \mathbb{R}^{M_{\mathbb{S}^2}}$ and a certain
amount of lens blurring characterized by axisymmetric convolution with a Gaussian smoothing
kernel with FWHM = $\Theta$. In this case observations are modeled by $\bm{y} = \bm{\mathsf{\Phi}}_{360^{\circ}} \bm{x} + \bm{n}$
for measurement operator
\begin{equation}
  \bm{\mathsf{\Phi}}_{360^{\circ}} = \bm{\mathsf{Y}}^{-1} \bm{\mathsf{\Theta}} \bm{\mathsf{Y}}
  \quad \text{and} \quad \bm{\mathsf{\Phi}}_{360^\circ}^\dagger = \bm{\mathsf{Y}}^\dagger
  \bm{\mathsf{\Theta}} \ (\bm{\mathsf{Y}}^{-1})^\dagger,
\end{equation}
where $\bm{\mathsf{Y}}, \bm{\mathsf{Y}}^{-1}$ are forward and inverse spherical harmonic transforms
correspondingly (see section \ref{sec:SphericalMaths}), $\bm{\mathsf{\Theta}}$ is the axisymmetric convolution with the harmonic
representation of $\Theta$.

As in the previous example the data-fidelity is given by the
$\snorm{\bm{\mathsf{\Phi}}_{360^{\circ}}\bm{x} - \bm{y}}_2^2$. Depending on the degree to which $\bm{x}$ is piece-wise
constant the TV-norm $\snorm{\bm{x}}_{\text{TV}} = \snorm{\nabla \bm{x}}_2$ (promoting gradient sparsity) constitutes a good choice of regularizer. For image deconvolution
an analysis wavelet sparsity promoting regularizer $\pnorm{\bm{\mathsf{\Psi}}^{-1} \bm{x}}_1$
is often also considered. Here we consider both regularization functionals $g(\bm{x}) = \lbrace \snorm{\bm{x}}_{\text{TV}}, \pnorm{\bm{\mathsf{\Psi}}^{-1} \bm{x}}_1\rbrace$
in the constrained analysis setting:
\begin{equation}
  \bm{x}^\star = \argminT_{\bm{x} \in \mathbb{C}^{N_{\mathbb{S}^2}}} \big \lbrace
  g(\bm{x})
  \big \rbrace \ \text{s.t.} \ \frac{1}{2\sigma^2}\snorm{\bm{\mathsf{\Phi}}_{360^{\circ}}\bm{x} - \bm{y}}_2^2 \leq \delta,
\end{equation}
where $\delta \in \mathbb{R}^+$ is the radius of the $\ell_2$-ball $\mathcal{B}_{\ell_2}^\delta$ which
balances sparsity against data-fidelity, and is defined straightforwardly from the known (in
general unknown) noise variance. We perform an example reconstruction with both priors in the
constrained formulation of the analysis problem using the proximal primal dual algorithm
\cite{Boyd2004Convex,Komodakis2015,Combettes2014}.
Both priors produce similar results, with wavelet sparsity regularization recovering SNR=$17.60$ dB and TV-norm marginally
superior at SNR = $17.65$ dB --- which can be seen in Figure \ref{fig:4pi_camera}. 

\subsection{CMB temperature anisotropies} \label{sec:cmb_temp}
Suppose one captures masked (and therefore incomplete) measurements of the cosmic microwave
background (CMB; \cite{Planck2018I}) $\bm{y} \in \mathbb{C}^{M_{\mathbb{S}^2}}$ that can be related through a
mapping operator $\bm{\mathsf{\Phi}}_{\text{CMB}} \in \mathbb{C}^{M_{\mathbb{S}^2} \times N_{\mathbb{S}^2}}$ to the
full-sky CMB signal $\bm{x} \in \mathbb{C}^{N_{\mathbb{S}^2}}$ which can be decomposed
into harmonic coefficients $\hat{x}_{\ell m} = \langle \bm{x}, Y_{\ell m} \rangle$ which for Gaussian fields such as CMB \cite{Planck2018I}) are uncorrelated and isotropic
$\mathbb{E} \lbrack \hat{x}^\ast_{\ell m} \hat{x}_{\ell^\prime m^\prime} \rbrack = \delta_{\ell \ell^\prime} \delta_{m m^\prime} C_{\ell}$, where ${C}_{\ell}$ is the angular power spectrum.

These considerations motivate the choice of a multivariate Gaussian prior $P(\bm{\hat{x}} | \mathcal{M}) = \exp \big ( -\bm{\hat{x}}^\dagger \bm{\mathsf{C^{-1}}}
  \bm{\hat{x}}/2 \big )$ for vectorized harmonic coefficients $\bm{\hat{x}}$ and covariance $\bm{\mathsf{C}}$ given by diagonal elements ${C}_{\ell}$.
Consider the case in which $\abs{M_{\mathbb{S}^2}} \ll \abs{N_{\mathbb{S}^2}}$ with
\textit{i.i.d.} Gaussian noise $\bm{n} \sim \mathcal{N}(0, \sigma^2) \in \mathbb{C}^{M_{\mathbb{S}^2}}$
then the whitened harmonic coefficients $\bm{\hat{x}}^\prime = \bm{\mathsf{C}^{-\frac{1}{2}}} \bm{\hat{x}}$ are modeled as $\bm{y} = \bm{\mathsf{\Phi}}_{\text{CMB}}\bm{\hat{x}}^\prime + \bm{n}$ for
measurement operator
\begin{equation}
  \bm{\mathsf{\Phi}}_{\text{CMB}} = \bm{\mathsf{D}} \bm{\mathsf{Y}} \bm{\mathsf{C}^{\frac{1}{2}}} \quad \text{and} \quad \bm{\mathsf{\Phi}}_{\text{CMB}}^\dagger =  \bm{\mathsf{C^{\frac{1}{2}}}} \bm{\mathsf{Y}}^\dagger \bm{\mathsf{D}}^\dagger ,
\end{equation}
for spherical harmonic transform $\bm{\mathsf{Y}}$ and masking and projection operators $\bm{\mathsf{D}},
  \bm{\mathsf{D}}^\dagger$ respectively. For diagonal noise covariance
$\Sigma = \sigma \mathbb{I}$ the univariate Gaussian likelihood is given by
$P(\bm{y} | \bm{\hat{x}}^\prime; \mathcal{M}) = \frac{1}{2\sigma^2} \snorm{\bm{\mathsf{\Phi}}_{\text{CMB}} \bm{\hat{x}}^\prime -
    \bm{y}}_2^2$, and so the synthesis unconstrained optimization is given by
\begin{align}
  \bm{\hat{x}}^{\text{MAP}\prime} & = \argminT_{\bm{\hat{x}}^\prime} \Big \lbrace
  \norm{\bm{\hat{x}}^\prime}_2^2 + \frac{1}{2\sigma^2} \snorm{\bm{\mathsf{\Phi}}_{\text{CMB}} \bm{\hat{x}}^\prime - \bm{y}}_2^2 \Big \rbrace,
\end{align}
where the pixel space signal is recovered by $\bm{x}^{\text{MAP}} = \bm{\mathsf{Y}}^{-1} \bm{\mathsf{C}^{\frac{1}{2}}} \bm{\hat{x}}^{\text{MAP}\prime}$.
This is the convex optimization formulation of what is commonly known as Wiener filtering,
which is often adopted for highly Gaussian fields, \emph{e.g.} the CMB. As expected for Wiener
filtering problems of this type (\textit{e.g.} Figure 6 in \cite{Kodi2019}), we recover
maps which exhibit inpainting of low-$\ell$ modes (large scale structure) into the masked
regions. The results of this experiment can be seen in Figure \ref{fig:CMB_example}.

\subsection{Weak gravitational lensing} \label{sec:WL}
The following example considers spherical imaging of dark matter.  
A more extensive analysis that applies the method presented to observational data, 
and not just simulations, is performed in \cite{M4} which leverages many of the methods developed in this work.
%
At first order, gravitational lensing manifests itself into the spin-$0$ convergence ${}_0\bm{x}(r, \omega) \in \mathbb{C}^{N_{\mathbb{S}^2}}$ (the integrated matter field along the line of sight) and the spin-$2$ shear ${}_2\bm{y}(
  r, \omega) \in \mathbb{C}^{N_{\mathbb{S}^2}}$ (the ellipticity of observed images) which can be related to the lensing 
  potential ${}_0\bm{\phi}(r, \omega) \in \mathbb{C}^{N_{\mathbb{S}^2}}$ by
$_0\bm{x}(r, \omega) = \frac{1}{4}(\eth_+ \eth_- + \eth_- \eth_+) \; {}_0\bm{\phi}$
and
$_2\bm{y}(r, \omega) = \frac{1}{2} \eth_+ \eth_+ \; {}_0\bm{\phi}$,
where $\eth_\pm$ are spin-raising/lowering operators \cite{Newman1966SpinSignals, Goldberg1967Harmonics}.
One can then relate $_0\bm{x}$ and $_2\bm{y}$ to one another in harmonic space by
${}_2\bm{\hat{y}}_{\ell m} = W_{\ell} \; {}_0\bm{\hat{x}}_{\ell m}$, for harmonic space kernel defined in
\cite{Wallis2017MASSMAPPY, M4}. 
As $_0\bm{x}$ is not directly observable, typically observations of $_2\bm{y}$ are collected
and used to reconstruct $_0\bm{x}$. Suppose one
recovers observations $\bm{y} \in \mathbb{C}^{M}$ which can be related to the
$\bm{x} \in \mathbb{C}^{N_{\mathbb{S}^2}}$ \textit{via} the forward model $\bm{\mathsf{\Phi}}_{\text{WL}} \in \mathbb{C}^{M \times N_{\mathbb{S}^2}} :\bm{x} \mapsto \bm{y}$.
Consider the scenario in which $\bm{y}$ are contaminated with \textit{i.i.d.} Gaussian noise
$\bm{n} \sim \mathcal{N}(0, \sigma^2) \in \mathbb{C}^{M}$,
then the observations are modeled by $\bm{y} = \bm{\mathsf{\Phi}}_{\text{WL}}\bm{x} + \bm{n}$ for
measurement operator
\begin{equation}
  \bm{\mathsf{\Phi}}_{\text{WL}} = \bm{\mathsf{D}} {}_2\bm{\mathsf{Y}}^{-1} \bm{\mathsf{W}} {}_0\bm{\mathsf{Y}}
  \ \ \text{and} \ \
  \bm{\mathsf{\Phi}}_{\text{WL}}^\dagger = {}_0\bm{\mathsf{Y}}^\dagger \bm{\mathsf{W}} ({}_2\bm{\mathsf{Y}}^{-1})^\dagger \bm{\mathsf{D}}^\dagger,
\end{equation}
for self-adjoint harmonic space multiplication $\bm{\mathsf{W}}$ with the axisymmetric kernel $W_{\ell}$,
masking and projection operators $\bm{\mathsf{D}}, \bm{\mathsf{D}}^\dagger$, and spin-$s$ forward and
inverse spherical harmonic transforms ${}_s\bm{\mathsf{Y}}, {}_s\bm{\mathsf{Y}}^{-1}$ respectively.

\begin{figure}

  \begin{tikzpicture}[spy using outlines={circle, magnification=4, connect spies}]
    \node[anchor=south west,inner sep=0] (image) at (0,0) {
      \begin{minipage}{.49\columnwidth}
        \includegraphics[width=\columnwidth, trim={10.6cm 11.1cm 10.6cm 5.6cm},clip]{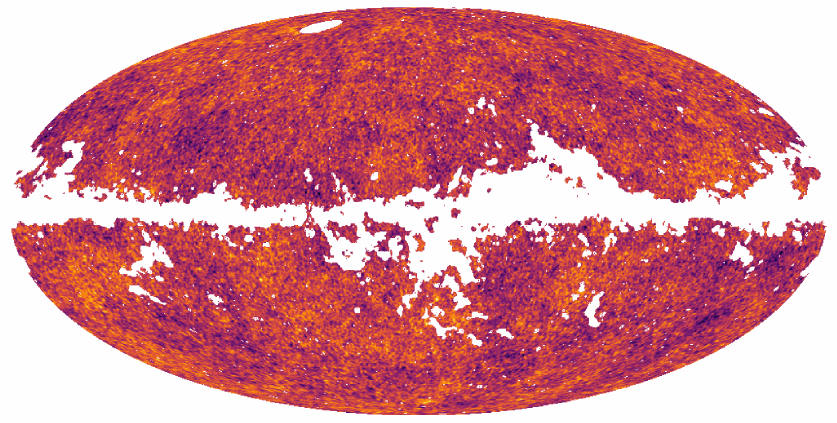}
        \put(-86, 65){Observations}
      \end{minipage}
      \begin{minipage}{.49\columnwidth}
        \includegraphics[width=\columnwidth, trim={10.6cm 5.6cm 10.6cm 11.1cm},clip]{cmb_example.pdf}
        \put(-90, 65){Reconstruction}
      \end{minipage}
    };

    \spy [black, size=1.5cm] on (5.2,0.90)
    in node[fill=white] at (7.7,0.1);

    \spy [black, size=1.5cm] on (0.75,0.90)
    in node[fill=white] at (3.2,0.1);

  \end{tikzpicture}%
  \vspace{-23pt}
  \caption{\textbf{Left:} Simulated Gaussian random field generated from lambda cold dark matter ($\Lambda$-CDM)
    best fit power spectrum, masked by a Planck survey mask \cite{Planck2018I} and polluted by
    30dB \textit{i.i.d.} Gaussian noise. \textbf{Right:} Unconstrained reconstruction using an $\ell_2$-norm
    Wiener prior solved by the proximal forward-backward algorithm, in the synthesis setting
    (for computational efficiency). The purpose of this reconstruction is to observe recovered low-$\ell$, large-scale information into the masked region (see \textit{e.g.} \cite{Kodi2019}).
  } \label{fig:CMB_example}
  \vspace{5pt}
  \begin{tikzpicture}[spy using outlines={circle, magnification=4, connect spies}]
    \node[anchor=south west,inner sep=0] (image) at (0,2.69+0.05) {
      \begin{minipage}{.49\columnwidth}
        \includegraphics[width=\columnwidth, trim={12.2cm 12.5cm 12.2cm 5.5cm},clip]{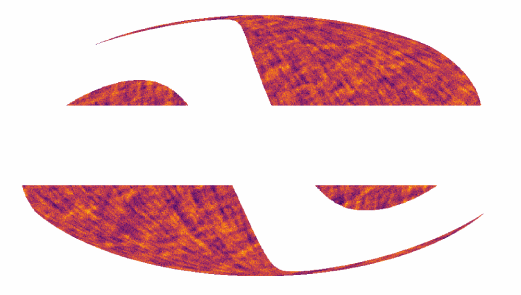}
        \put(-95, 73){Real observations}
      \end{minipage}
      \begin{minipage}{.49\columnwidth}
        \includegraphics[width=\columnwidth, trim={12.2cm 9.0cm 12.2cm 9.1cm},clip]{weak_lensing_example.pdf}
        \put(-95, 73){Imag observations}
      \end{minipage}
    };
    \node[anchor=south west,inner sep=0] (image_2) at (0,0) {
      \hspace{0.25\columnwidth}
      \begin{minipage}{.49\columnwidth}
        \includegraphics[width=\columnwidth, trim={12.2cm 5.0cm 12.2cm 12.7cm},clip]{weak_lensing_example.pdf}
        \put(-90, 79){Reconstruction}
      \end{minipage}
    };

    \spy [black, size=1.5cm] on (2.5,4.54)
    in node[fill=white] at (3.4, 3.7+0.1);

    \spy [black, size=1.5cm] on (6.95,4.5)
    in node[fill=white] at (3.4+4.45 ,3.7+0.1);

    \spy [black, size=1.5cm] on (4.85,2.17)
    in node[fill=white] at (3.4+2.4, 0.9+0.4);

  \end{tikzpicture}%
  \vspace{-25pt}
  \caption{\textbf{Top:} Simulated weak lensing shear field generated from a ground
    truth N-body simulation \cite{Takahasi2017} signal, which was further contaminated with
    5dB \textit{i.i.d.} Gaussian noise and masked using realistic pseudo-Euclid survey mask.
    \textbf{Bottom:} Unconstrained reconstructed dark matter mass-map using $\ell_1$-norm wavelet sparsity
    prior solved through using the proximal forward-backward algorithm in the analysis setting.
    See related works \cite{M4} for a comprehensive analysis of this application, with uncertainty quantification.}
  \label{fig:Lensing}
\end{figure}

Since principled statistical interpretation is crucial for this science application,  one may consider the unconstrained formulation of this inverse problem which we solve here with the
proximal forward-backward algorithm in the analysis setting, with univariate Gaussian
likelihood (data-fidelity) and sparsity promoting Laplace type spherical wavelet prior
(regularizer),
\begin{equation}
  \bm{x}^{\text{MAP}} = \argminT_{\bm{x} \in \mathbb{C}^{N_{\mathbb{S}^2}}} \Big \lbrace
  \lambda \pnorm{\bm{\mathsf{\Psi}}^{-1}\bm{x}}_1 + \frac{1}{2\sigma^2}\snorm{\bm{\mathsf{\Phi}}_{\text{WL}}\bm{x} - \bm{y}}_2^2 \Big \rbrace,
\end{equation}
for automatically marginalized regularization parameter $\lambda \in \mathbb{R}^+$ (see Section
\ref{sec:optimisation_setting}). Note that sparse priors are often adopted in the weak lensing
setting, recovering state-of-the-art results \cite{Lanusse2016, M1, M4}.
Images from this experiment can be seen in Figure \ref{fig:Lensing}. A more in depth application of the methods developed in this paper to dark matter reconstruction can be found in
\cite{M4}, in which global hypothesis testing (leveraging the techniques of Section~\ref{sec:hypothesis}) is performed to determine whether two reconstruction methods produce commensurate estimates.

\section{Conclusion} \label{sec:conclusions}

We present and discuss a flexible, general framework for variational imaging on the
sphere. We consider different formulations of inverse problems as either constrained or unconstrained problems \cite{Boyd2004Convex}
in both the analysis and synthesis settings \cite{Elad2006, Wallis2017Sparse}. The implications,
advantages, and disadvantages of each choice within the context of imaging on the sphere is
considered both qualitatively and quantitatively.
Crucially, we highlight the direct relationship between the unconstrained setting and
Bayesian inference. We combine this realization with recent developments in the field of
probability density theory \cite{Pereyra2017MAPUQ} to demonstrate how one can perform
rapid, statistically principled uncertainty quantification on reconstructed signals (building
upon work in \cite{Pereyra2017MAPUQ, Cai2018UQ, Repetti2018, M1, M2, M3}). Furthermore, we demonstrate mathematically how one may exploit
linearity and general inequality relations to dramatically accelerate such uncertainty
quantification techniques in all settings.  It is shown that in a variety of interesting cases
these uncertainty quantification techniques reduce to computationally trivial 1-dimensional $P^{\text{th}}$-order polynomial
root finding problems, which can often be solved analytically. While such computational savings
are key for scalable, statistically principled spherical imaging, they are likely also of use for 
standard 2-dimensional Euclidean imaging.

The aforementioned techniques are demonstrated on an extensive suite of numerical
experiments, which simulate a diverse set of typical use-cases. Specifically, we consider
a spread of deconvolution, inpainting, and de-noising problems, \emph{e.g.} from resolving blurred $360^{\circ}$ camera images, to imaging
the dark matter distribution on the celestial sphere. It is found that that optimality of
problem formulation (constrained \emph{versus} unconstrained) and setting (analysis
\emph{versus} synthesis) is highly situationally dependent on the sphere. The authors 
make the scalable, open-source spherical reconstruction software developed during this work
(S2INV), publicly available.

\section*{Acknowledgment}

MAP is supported by the Science and Technology Facilities Council (STFC).
This work was also supported in part by the Leverhulme Trust. LP is supported by the Dunlap Institute,
which is funded through an endowment established by the David Dunlap family and the University
of Toronto.

\ifCLASSOPTIONcaptionsoff
  \newpage
\fi



\vfill
\end{document}